\documentstyle[epsf,12pt]{article}
\newcommand {\beq}{\begin{equation}}
\newcommand {\eeq}{\end{equation}}
\newcommand {\beqa}{\begin{eqnarray}}
\newcommand {\eeqa}{\end{eqnarray}}
\newcommand {\n}{\nonumber \\}

\renewcommand{\theequation}{\thesection.\arabic{equation}}
\begin{document}
\setlength{\oddsidemargin}{0cm}
\setlength{\baselineskip}{7mm}

\begin{titlepage}
 \renewcommand{\thefootnote}{\fnsymbol{footnote}}
$\mbox{ }$
\begin{flushright}
\begin{tabular}{l}
KEK-TH-1020\\
May  2005
\end{tabular}
\end{flushright}

~~\\
~~\\
~~\\

\vspace*{0cm}
    \begin{Large}
       \vspace{2cm}
       \begin{center}
{Stability of Fuzzy $S^2 \times S^2 \times S^2$\\ 
in IIB Type Matrix Models}
\\
       \end{center}
    \end{Large}

  \vspace{1cm}

\begin{center}
           Hiromichi K{\sc aneko}$^{2)}$\footnote
           {
e-mail address : kanekoh@post.kek.jp},
            Yoshihisa K{\sc itazawa}$^{1),2)}$\footnote
           {
e-mail address : kitazawa@post.kek.jp}\\{\sc and}
           Dan T{\sc omino}$^{1)}$\footnote
           {
e-mail address : dan@post.kek.jp}

        $^{1)}$ {\it High Energy Accelerator Research Organization (KEK),}\\
               {\it Tsukuba, Ibaraki 305-0801, Japan} \\
        $^{2)}$ {\it Department of Particle and Nuclear Physics,}\\
                {\it The Graduate University for Advanced Studies,}\\
{\it Tsukuba, Ibaraki 305-0801, Japan}\\
\end{center}

\vfill

\begin{abstract}
\noindent
We study the stability of fuzzy $S^2 \times S^2 \times S^2$ backgrounds in 
three different IIB type matrix models with respect to the change of the spins 
of each $S^2$ at the two loop level.
We find that $S^2 \times S^2 \times S^2$ background is metastable and the 
effective action  favors a single large $S^2$ in comparison to the remaining 
$S^2 \times S^2$
in the models with Myers term.
On the other hand, we find that a large $S^2 \times S^2$ in comparison to
the remaining $S^2$  is favored in IIB matrix model itself.
We further study the stability of fuzzy $S^2 \times S^2$ background in detail
in IIB matrix model with respect to the scale factors
of each $S^2$ as well. 
In this case, we find unstable directions which lower the effective action away from the
most symmetric fuzzy $S^2 \times S^2$ background.
\end{abstract}
\vfill
\end{titlepage}
\vfil\eject

\section{Introduction}
\setcounter{equation}{0}
\setcounter{footnote}{0}

Why the spacetime is four dimensional is an interesting question in our quest to 
understand the origin of universe.
Nowadays sting theory is a leading candidate of quantum gravity, 
and IIB matrix model is a candidate of a non-perturbative  formulation of string theory 
\cite{IKKT96, FKKT98}.
Matrix models may be useful to answer this question if they can consistently describe
the dynamics of gauge fields and quantum gravity as expected.

In IIB matrix model, the spacetime may be  represented by a particular configuration of matrices.
When we are given a matrix configuration as a background field, 
we can examine its stability by investigating the behavior of the effective action  under the
change of some parameters of the background. 
In this framework, the stabilities of fuzzy $S^2$, $S^2 \times S^2$ \cite{IKTT03,IT03} and 
$T^2\times T^2$ \cite{BHKK04} have been studied.
This paper extends these investigations into a simple 6 dimensional manifold.

In the previous investigations, the large $N$ scaling behavior of the NC gauge theory
on these manifolds has been clarified \cite {IKTT03}.  In supersymmetric models,
the effective action scales as $N^2$, $N$ and
$N^{4\over 3}$ in 2, 4 and 6 dimensional manifolds respectively
\footnote{We subtract the universal gauge volume of $SU(N)/Z_N$ from the matrix model 
effective action.}. 
It always scales as $N^2$ and the one loop approximation is exact in bosonic models
\cite{ABNN04}.
A major purpose of this paper is to explicitly
verify the predicted scaling behavior of the effective action 
$(N^{4\over 3})$ at the two loop level for a simple 6d fuzzy manifold.

In Gaussian approximations, it has been found that 4 dimensional spacetime
tends to minimize the effective action \cite{NS, Kyoto}.
The advantage of our approach is that the effective actions on homogeneous spaces
are much lower than those in the Gaussian approximations which are $O(N^2)$.
It is because supersymmetry is broken only softly on homogeneous spaces.
Although IIB matrix model is supersymmetric,  supersymmetry cannot 
be respected on compact homogenous spacetime.
It is precisely why we obtain nonvanishing effective action on these manifolds.
It has been a great challenge to explain 4 dimensionality in string theory
since the vacuum degenerates as long as it is supersymmetric.
In nonperturbative formulation of string theory, the extension of (Euclidean) spacetime
may be inevitably finite. Such a view is consistent with finite entropy of
de Sitter spacetime in which we are likely to live in.
IIB matrix model suggests that the vacuum 
degeneracy may be lifted due to finite extension of spacetime.

Fuzzy $S^2$ can be embedded in Hermitian matrices as
\beq
A_{i}=fj_i ,
\label{S2}
\eeq
where $j_i$ are the generators of $SU(2)$ with spin $l$ and $f$ denotes the scale factor
of this background.
The spin $l$ determines the size of $S^2$ in our group theoretic construction
of $S^2=SU(2)/U(1)$
and we can also introduce the scale factor $f$ to vary the overall size of $S^2$.
Thus the parameters of $S^2$ in matrix models are a spin $l$ and 
a scale factor $f$ for each $S^2$. 

After setting up our calculation procedure in section 2,
we first investigate the stabilities of $S^2 \times S^2 \times S^2$ in three different matrix models 
with respect to the variation of the spins while assuming the identical scale factor $f$
for all $S^2$'s.
In section 3 and 4
we find that $S^2 \times S^2 \times S^2$ background is metastable and the 
effective action  favors a single large $S^2$ in comparison to the remaining $S^2 \times S^2$
in the matrix models with Myers term.
On the other hand, we find that a large $S^2 \times S^2$ in comparison to
the remaining $S^2$  is favored in IIB matrix model itself in section 5.
These findings are consistent with previous studies \cite{IKTT03,IT03}.
We subsequently investigate $S^2 \times S^2$ and $S^2 \times S^2 \times S^2$ in 
IIB matrix model with respect to the variations of the spins and scale factors in section 6 and 7.
In this case, we find unstable directions which lower the effective action away from the
most symmetric fuzzy $S^2 \times S^2$ background as suggested by \cite{BHKK04}.
We conclude in section 8 with discussions.

\section{IIB type matrix models on fuzzy $S^2 \times S^2 \times S^2 $ }
\setcounter{equation}{0}

Since our motivation is to explain the 4 dimensionality of spacetime in superstring theory,
it is natural to investigate the stability of fuzzy $S^2 \times S^2 \times S^2 $
in IIB matrix model \cite{IKKT96}
\begin{equation} \label{action_IIB}
S_{IIB} = - \frac{1}{4} Tr \left[ A_{\mu} , A_{\nu} \right]^2 - \frac{1}{2} 
Tr \bar{\psi} \Gamma_\mu \left[ A_\mu , \psi \right] ,
\end{equation}
where $A_{\mu}$ and $\psi$ are $N \times N$ Hermitian matrices. 
We also investigate a deformed model with Myers term\cite{My99,IKTW,Ki02,Bone}
\begin{equation} \label{action_Myers}
S_{Myers} = \frac{i}{3} f_{\mu \nu \rho} Tr \left[ A_\mu ,A_\nu \right] A_\rho.
\end{equation}
It is because fuzzy $S^2$, $S^2\times S^2$ and $S^2 \times S^2 \times S^2 $ 
become  classical solutions after such a deformation since
the equations of motion are
\beqa
\left[ A_\mu , \left[ A_\mu , A_\nu \right] \right] + \{\bar{\psi}, \Gamma_\nu \psi \}
&=& \bigg\{ 
\begin{array}{cc}
 - i f_{\mu\nu\rho} \left[ A_\mu , A_\rho \right] & \mathrm{(with ~Myers ~term)} \n
 0 & \mathrm{(without~ Myers ~term)}
\end{array} , \n
\Gamma_\mu \left[ A_\mu ,\psi \right] &=& 0 .
\eeqa
This modification does not alter the convergence properties of IIB matrix model
\cite{AW} while it breaks supersymmetry softly.

After adopting fuzzy $S^2 \times S^2 \times S^2 $ as a background field, we separate 
$A_{\mu}$ and $\psi$ into background fields $p_\mu$, $\chi$ and quantum fluctuations 
$a_\mu$, $\varphi$.
\beqa \label{sep_GF}
A_\mu &=& p_\mu + a_\mu , \n
\psi &=& \chi + \varphi .
\eeqa
Since fuzzy $S^2$ can be realized as in (\ref{S2}), 
we take the following $p_{\mu}$ and $\chi$ to represent fuzzy $S^2 \times S^2 \times S^2 $ as
\beqa
\label{classical-sol}
p_{\mu} &=& f \left( \bar{j}_{\mu} \otimes 1 \otimes 1 \right) \otimes 1_n \qquad  
( \mu = 1,2,3 ) , \n
p_{\mu} &=& f \left( 1 \otimes \hat{j}_{\mu} \otimes 1 \right) \otimes 1_n \qquad  
( \mu = 4,5,6 ) , \n
p_{\mu} &=& f\left( 1 \otimes 1 \otimes \tilde{j}_{\mu} \right) \otimes 1_n \qquad  
( \mu = 7,8,9 ) , \n
p_{0} &=& 0 , \n
\chi &=& 0.
\eeqa
where $\bar{j}_{\mu}, \ \hat{j}_{\mu}$ and $\tilde{j}_{\mu}$ are the angular momentum
operators with spin 
$l_1$, $l_2$ and $l_3$ respectively.
$1_n$ represents the $n\times n$ identity  corresponding to $n$ coincident fuzzy 
$S^2 \times S^2 \times S^2 $. 
This background is a classical solution when Myers term is added while
it becomes a quantum solution in IIB matrix model at the two loop level
when the effective action is stationary with respect to $f$.

$p_\mu$'s satisfy the $SU(2)$ algebras as follows
\beqa
[p_{\mu},p_{\nu}] &=& i f_{\mu \nu \rho} p_{\rho} , \n
f_{\mu \nu \rho} &=& \left\{ 
\begin{array}{cc}
f \epsilon_{\mu \nu \rho} & (\mu ,\nu ,\rho ) \in (1,2,3) \n
f \epsilon_{\mu \nu \rho} & (\mu ,\nu ,\rho ) \in (4,5,6) \n
f \epsilon_{\mu \nu \rho} & (\mu ,\nu ,\rho ) \in (7,8,9) \n
0 & \mathrm{(others)}
\end{array}
\right. .
\eeqa
We can construct the Casimir operators in the respective representations as
\beqa
\sum_{\mu=1,2,3} (p_{\mu})^2 &=& f^2 l_1(l_1+1) , \n
\sum_{\mu=4,5,6} (p_{\mu})^2 &=& f^2 l_2(l_2+1) , \n
\sum_{\mu=7,8,9} (p_{\mu})^2 &=& f^2 l_3(l_3+1) .
\eeqa
Since the right hand side of the above equations are the squared radii,
the spin and scale factor $f$  determine the size of each $S^2$.
The dimension of the matrices is
\begin{equation}
N=n(2l_1+1)(2l_2+1)(2l_3+1) .
\end{equation}

As we can see, this background represents a simple six dimensional spacetime.
The choice of $S^2$ facilitates us to carry out detailed analysis of the effective action since
the representations of $SU(2)$ are well known.
By varying the representations (spins) 
of respective $S^2$, we can explore the whole range of
the manifolds from the 2 dimensional limit with a single large $S^2$ to the 6 dimensional limit
with a large $S^2\times S^2 \times S^2 $ with the identical radii.
We can thus explore which dimensionality between 2 and 6 
is favored by the effective action in this class of backgrounds.

The effective action can be evaluated in a background gauge method by substituting (\ref{sep_GF}) 
for (\ref{action_IIB}) and (\ref{action_Myers}).
 We introduce a gauge fixing term $S_{gf}$ and a ghost term $S_{gh}$ for gauge fixing. 
 The gauge fixed action is
\beqa \label{q_action_IIB}
S'_{IIB} &\equiv& S_{IIB} + S_{gh} + S_{gf} \n
&=& - \frac{1}{4} Tr [p_{\mu},p_{\nu}]^2- Tr a_{\rho} ([p_{\mu},[p_{\rho},p_{\mu}]]) \n
&&+ \frac{1}{2} Tr  a_{\mu} ( \delta^{\mu \nu} P^2 + 2if_{\mu \nu \rho} P_{\rho} ) a_{\nu} - \frac{1}{2} 
Tr \bar{\varphi} \Gamma^{\mu} P_{\mu} \varphi + Tr b P^2 c \n
&&- Tr P_{\mu} a_{\nu} \left[ a_{\mu}, a_{\nu} \right] - \frac{1}{4} Tr \left[ a_{\mu} , a_{\nu} \right]^2 \n
&&- \frac{1}{2} Tr \bar{\varphi} \Gamma_\mu \left[ a_\mu , \varphi \right] + 
Tr b P_{\mu}\left[  a_{\mu} ,c \right]  ,
\eeqa
\beqa
\label{q_action_Myers}
S_{Myers} &=& \frac{i}{3} f_{\mu \nu \rho} Tr \left[ p_\mu ,p_\nu \right] p_\rho + i f_{\mu \nu \rho} 
Tr \left[ p_\mu ,p_\nu \right] a_\rho \n
&&- i f_{\mu \nu \rho} Tr a_{\mu} P_{\rho} a_{\nu}+ 
\frac{i}{3} f_{\mu \nu \rho} Tr \left[ a_\mu ,a_\nu \right] a_\rho .
\eeqa
where $P_{\mu} X = [ p_{\mu} , X] $, and
\beqa
S_{gh} &=& Tr bP_{\mu} [p_{\mu}+a_{\mu},c]  ,\n
S_{gf} &=& - \frac{1}{2} Tr(P_{\mu}a_{\mu})^2.
\eeqa

\section{Bosonic  matrix model with Myers term}
\setcounter{equation}{0}

In this section, we investigate a bosonic matrix model with Myers term
whose action is
\begin{equation} \label{boson-action}
S = - \frac{1}{4} Tr \left[ A_{\mu} , A_{\nu} \right]^2 + 
\frac{i}{3} f_{\mu \nu \rho} Tr \left[ A_\mu ,A_\nu \right] A_\rho ,
\end{equation}
where $\mu =0,1,\cdots, 9$ in correspondence to IIB matrix model.
We certainly require 9 matrices to construct fuzzy $S^2\times S^2\times S^2$.
Although we are eventually interested in IIB matrix model, bosonic models
are simple enough to admit exact solutions in the large $N$ limit.
Our prediction can be verified by comparing with numerical investigations
\cite{ABNN04} .
For this purpose,
it is useful to generalize the classical solutions as follows:
\beqa
p_{\mu} &=& \beta \left( \bar{j}_{\mu} \otimes 1 \otimes 1 \right) 
\otimes 1_n \qquad  ( \mu = 1,2,3 ) , \n
p_{\mu} &=& \beta \left( 1 \otimes \hat{j}_{\mu} \otimes 1 \right) 
\otimes 1_n \qquad  ( \mu = 4,5,6 ) , \n
p_{\mu} &=& \beta \left( 1 \otimes 1 \otimes \tilde{j}_{\mu} \right) 
\otimes 1_n \qquad  ( \mu = 7,8,9 ) , \n
p_{0} &=& 0 , \n
\chi &=& 0 .
\eeqa
The shift of $\beta$ away from the classical value of $f$ is required to take account of
quantum effects.  Since the tree and the one loop contributions dominate in bosonic
models in the large $N$ limit as we shall see, this approach enables us an exact investigation
within this class of the backgrounds.

The tree level effective action is
\beqa
\Gamma_{tree} &=& - \frac{1}{4} Tr [p_{\mu},p_{\nu}]^2 + 
\frac{i}{3} f_{\mu \nu \rho} Tr \left[ p_\mu ,p_\nu \right] p_\rho \n
&\rightarrow& 8 N^{2} \left[ \frac{1}{2} \left( \frac{\beta}{f} \right)^4 - 
\frac{2}{3} \left( \frac{\beta}{f} \right)^3 \right]  
\frac{ f^4 }{ 2^5 n^{\frac{2}{3}} N^{\frac{1}{3}} } 
(r_1 + r_2 + r_3 ) .
\eeqa
Here the large $N$ limit is taken when we proceed from the 1st to 2nd line
($l_1,l_2,l_3 \gg 1$) and the following
ratios which measure the relative sizes of $S^2$'s are introduced
\begin{equation}
r_1=\left( \frac{l_1 l_1}{l_2 l_3} \right)^{\frac{2}{3}} \ , 
\ r_2=\left( \frac{l_2 l_2}{l_1 l_3} \right)^{\frac{2}{3}} \ , 
\ r_3=\left( \frac{l_3 l_3}{l_1 l_2} \right)^{\frac{2}{3}} .
\end{equation}
The leading term of the one loop effective action in the large $N$ limit can be evaluated as
\beqa
\Gamma_{1-loop} &=& \frac{1}{2} Tr \log (P^2 \delta_{\mu \nu}) -Tr log P^2 \n
&\rightarrow& 8 N^2 \Bigl( \log \beta + \frac{1}{3} \log \frac{N}{8n} \n
&&+ {1\over 128}\int_0^4 dX \, dY \, dZ \log (r_1 X + r_2 Y + r_3 Z ) \Bigr) .
\eeqa

The magnitude of the two loop effective action can be estimated by 
the following power counting argument.  
The two loop amplitude diverges as the eighth power of the cutoff 
($N^{\frac{8}{3}}$) while it is suppressed by a power of $N$
which is associated with the interaction vertices.
Since the one loop amplitude is $O(N^2)$, we need to adopt the 
 't Hoot coupling ($N^{1\over 3}/f^4$) in such a way to make the tree contribution 
 of $O(N^2)$ as well.
 With such a choice, the two loop amplitude scales as $N^{\frac{4}{3}}$.
 Since the tree and the one loop amplitude scale as $N^2$, we can safely ignore
 the two loop amplitude in the large $N$ limit.
 The analogous argument shows that we can ignore all higher loop contributions and
 the one loop effective action becomes exact in the large $N$ limit.

In this way, we obtain the effective action in the large $N$ limit as
\beqa\label{boson-effact}
\Gamma &=& 8 N^2 \Biggl( \left[ \frac{1}{2} \left( \frac{\beta}{f} \right)^4 - \frac{2}{3} 
\left( \frac{\beta}{f} \right)^3 \right]  \frac{ 1 }{ 2^5 \lambda_1^2 } (r_1 + r_2 + r_3 ) \n
&&+  \log \beta + \frac{1}{3} \log \frac{N}{8n} + {1\over 128}\int_0^4 dX \, dY \, dZ 
\log (r_1 X + r_2 Y + r_3 Z )	 \Biggr) ,
\eeqa
where $\lambda_1$ is the 't Hooft coupling:
\begin{equation}
\lambda_1^2 =  \frac{ n^{\frac{2}{3}} N^{\frac{1}{3}} }{ f^4 } .
\end{equation}

To minimize the effective action, we have to solve $\frac{\partial \Gamma}{\partial \beta}=0$. 
This condition determines the scale factor $\beta$ as
\beqa
\label{boson-sol-beta}
\frac{\beta}{f} &=& \frac{1}{4} + \frac{1}{2} \sqrt{\frac{1}{4}+g(A)} +
\frac{1}{2} \sqrt{ \frac{1}{2}-g(A) +\frac{1}{4\sqrt{\frac{1}{4}+g(A)}} } ,
\eeqa
where
\beqa
g(A) &=& \frac{ 2^{\frac{5}{3}} A }{ 3^{\frac{1}{3}} 
\left( 9A+\sqrt{3}\sqrt{27A^2-128A^3} \right)^{\frac{1}{3}} } + 
\frac{ \left( 9A+\sqrt{3}\sqrt{27A^2-128A^3} \right)^{\frac{1}{3}} }{ 6^{\frac{2}{3}} } ,
\eeqa
and
$A =2^5 \lambda_1^2$.
We find that this solution exists in the weak coupling regime where
\begin{equation}
\lambda_1^2 = \frac{ n^{\frac{2}{3}} N^{\frac{1}{3}} }{ f^4 } < 
\frac{3^3}{2^{12}}(r_1 + r_2 + r_3) \simeq 0.0198~~ {\mathrm at} ~~r_1=r_2=r_3=1.
\end{equation}
Beyond this critical point, the fuzzy $S^2\times S^2\times S^2$ background no longer exists. 
Just like  \cite{ABNN04}, this point separates
the background distributions between the collapsed phase and the fuzzy 
$S^2\times S^2\times S^2$ phase.

\begin{figure}[htbp]
\epsfysize=8cm
\begin{center}
\vspace{1cm}
\hspace{0cm}
\epsfbox{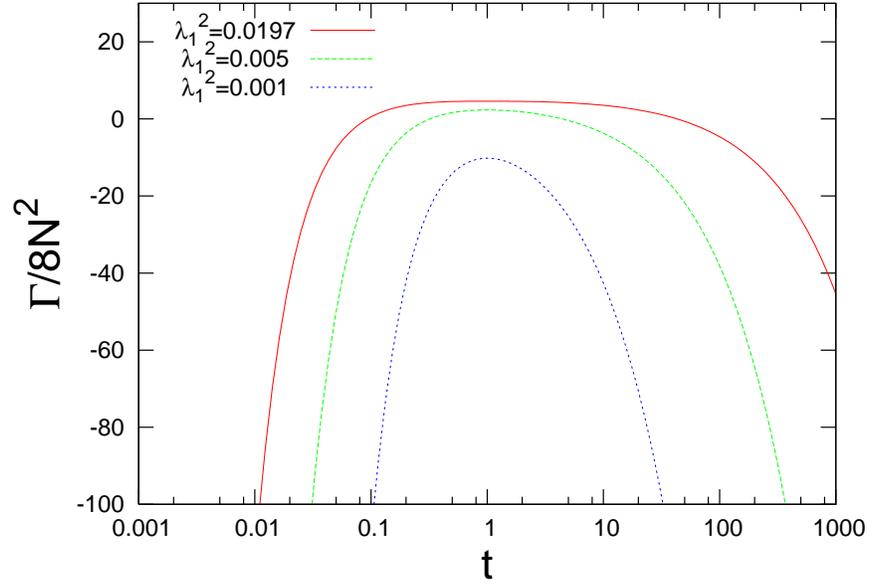}
\end{center}
\caption{${\Gamma}/{8N^2}$ against $t={l_1}/{l_3}={l_2}/{l_3}$ 
for $N=8 \times 10^6$ and $\lambda_1^2=0.0197,0.005,0.001$.
}
\label{fig:eff_bos_2d}
\end{figure}

\begin{figure}[htbp]
\epsfysize=8cm
\begin{center}
\vspace{1cm}
\hspace{0cm}
\epsfbox{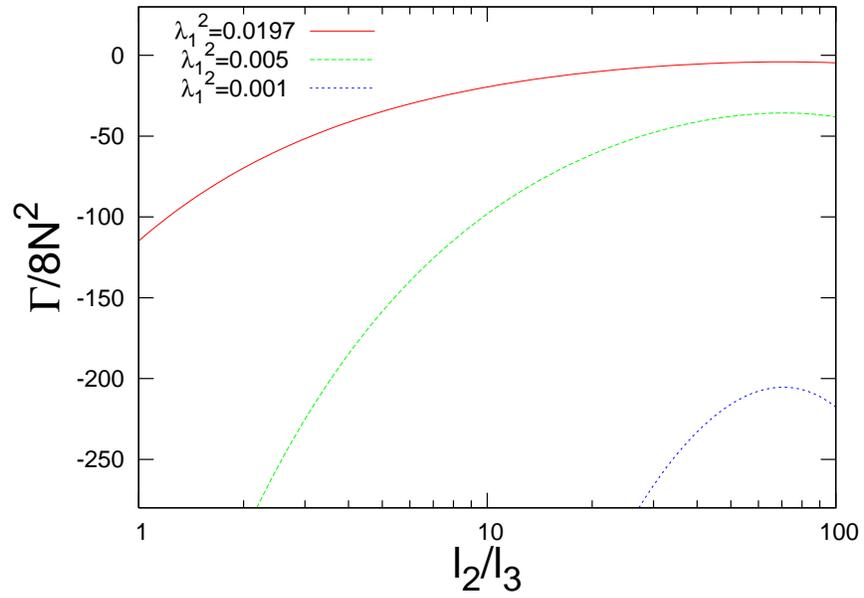}
\end{center}
\caption{${\Gamma}/{8N^2}$ against ${l_2}/{l_3}$
for ${l_1}/{l_3}=100$ , $N=8 \times 10^6$ and
$\lambda_1^2=0.0197,0.005,0.001$.}
\label{fig:eff_bos_2d2}
\end{figure}

After plugging (\ref{boson-effact}) into (\ref{boson-sol-beta}), we can estimate
the effective action by performing the triple integrals numerically for each $\lambda_1$. 
Fig.\ref{fig:eff_bos_2d} shows ${\Gamma}/{8N^2}$ against $t={l_1}/{l_3}={l_2}/{l_3}$ 
for $N=8 \times 10^6$ and $\lambda_1^2=0.0197,0.005,0.001$. 
$t=1$ corresponds to the most symmetric fuzzy $S^2 \times S^2 \times S^2$ background.
As $t\rightarrow 
\infty$,  the background approaches fuzzy $S^2 \times S^2$ while
it  approaches fuzzy $S^2$ as $t\rightarrow 0$.

From  Fig.\ref{fig:eff_bos_2d} we can see that fuzzy $S^2 \times S^2 \times S^2 $ background 
is not stable. 
To explore whether it tends to decay into $S^2 \times S^2$ or $S^2$, let us examine the behavior 
of the effective action in more details.
When we fix ${l_1}/{l_3}=100$ and vary ${l_2}/{l_3}$ between
1 to 100, the action behaves as Fig.\ref{fig:eff_bos_2d2}.
Here, ${l_2}/{l_3}=100$ and $\ {l_2}/{l_3}=1$ in Fig.\ref{fig:eff_bos_2d2}
correspond to $t=100$ and $t=0.01$ in Fig.\ref{fig:eff_bos_2d}
respectively because of the equivalence of the three $S^2$'s.
As we observe $\Gamma$ is smooth with respect to ${l_1}/{l_3}$ and ${l_2}/{l_3}$, 
and does not develop a local minimum, 
we can convince ourselves that $S^2$ is favored in this example.
The situation like this holds for the other combinations of ${l_1}/{l_3}$ and ${l_2}/{l_3}$ 
when one of them is large enough.
Therefore we conclude that fuzzy $S^2 \times S^2 \times S^2 $ background is not stable in the 
action (\ref{boson-action}), and it decays toward fuzzy $S^2$.

\section{Deformed IIB matrix model with Myers term}
\setcounter{equation}{0}

In this section, we study a deformed IIB matrix model with Myers term whose action is
\begin{equation} \label{Myers-deform-action}
S = - \frac{1}{4} Tr \left[ A_{\mu} , A_{\nu} \right]^2  
-  \frac{1}{2} Tr \bar{\psi} \Gamma_\mu \left[ A_\mu , \psi \right] 
+ \frac{i}{3} f_{\mu \nu \rho} Tr \left[ A_\mu ,A_\nu \right] A_\rho.
\end{equation}
We investigate the stability of the fuzzy $S^2 \times S^2 \times S^2 $ background 
(\ref{classical-sol}) which is a classical solution in this action 
\footnote{We need not generalize $f$ into $\beta$ here 
since there are no tadpoles at the one loop level .}.

The effective action can be evaluated by extending the procedures in \cite{IKTT03}. 
The results in the large N limit are
\beqa
\Gamma_{tree} &\simeq& - n^{\frac{2}{3}} N^{\frac{4}{3}} \frac{1}{24} 
\frac{ f^4 N^{\frac{1}{3}} }{ n^{\frac{4}{3}} } (r_1 + r_2 + r_3 ), \n
\Gamma_{1-loop} &\simeq& n^{\frac{2}{3}} N^{\frac{4}{3}} \frac{1}{8} 
 \int_0^4 dX \int_0^4 dL \int_0^4 dA \frac{1}{r_1 X+r_2 L+r_3 A} , \n
\Gamma_{2-loop} &\simeq& - n^{\frac{2}{3}} N^{\frac{4}{3}} 
\frac{ n^{\frac{4}{3}} }{ f^4 N^{\frac{1}{3}} } \left( \frac{3}{2} f_3 + 4 f_4 \right),
\eeqa
$f_3$ and $f_4$ are defined as the following multiple integrals
\beqa
f_3&=& \int^4_0 \frac{dXdYdZdLdMdNdAdBdC}
{(r_1 X + r_2 L+ r_3 A)( r_1 Y + r_2 M+ r_3 B)(r_1 Z + r_2 N+ r_3 C)} \n
&&\times  W(X,Y,Z) W(L,M,N) W(A,B,C) , \n
f_4&=& \int^4_0 \frac{dXdYdZdLdMdNdAdBdC}
{(r_1 X + r_2 L+ r_3 A)^2 ( r_1 Y + r_2 M+ r_3 B)^2 (r_1 Z + r_2 N+ r_3 C)} \n
&&\times \Bigl[ r_1^2 XY + r_2^2 LM + r_3^2 AB +\frac{1}{2r_3} (Z-X-Y) (N-L-M) \n
&&\times +\frac{1}{2r_2} (Z-X-Y) (C-A-B) +\frac{1}{2r_1} (N-L-M) (C-A-B) \Bigr] \n
&&\times  W(X,Y,Z) W(L,M,N) W(A,B,C) ,
\eeqa
where
$W(X,Y,Z)$ is the asymptotic formula of Wigner's $6j$ symbols 
which appear in the interaction vertices:
\beqa
l_1^3 \left\{ \begin{array}{ccc} j_1 & j_2 & j_3 \\ l_1 & l_1 & l_1 \end{array} \right\}^2 
&\simeq& W(j_1^2,j_2^2,j_3^2)\n
&=&\frac{1}{2 \pi \sqrt{ \frac{YZ(4-Y)(4-Z)}{4} -
 \left( X-\frac{2Y+2Z-YZ}{2} \right)^2 } } .
\eeqa
This approximation is valid in the uniformly large angular momentum regime.
Since the effective action is highly divergent, we argue that 
this approximation is exact in the large $N$ limit.

\begin{figure}[htbp]
\epsfysize=8cm
\begin{center}
\vspace{1cm}
\hspace{0cm}
\epsfbox{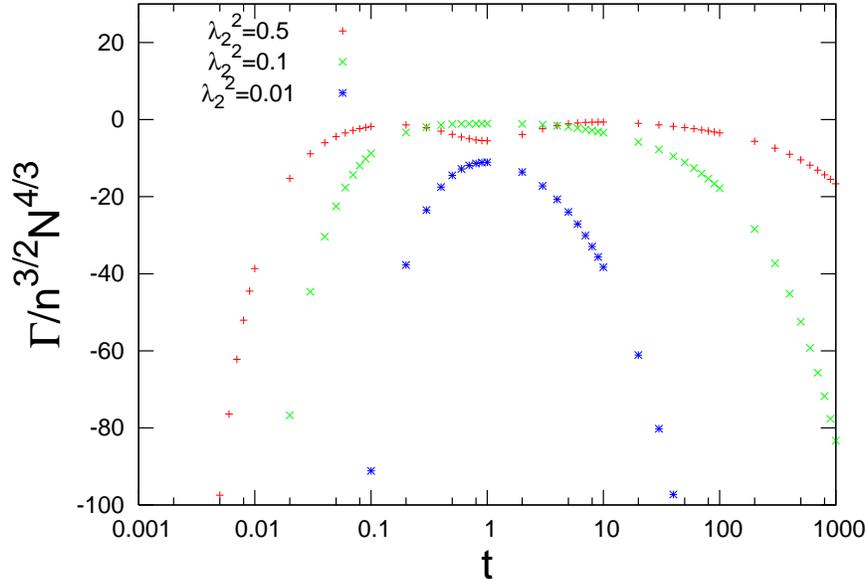}
\end{center}
\caption{${\Gamma}/{n^{3/2}N^{4/3}}$ against $t={l_1}/{l_3}={l_2}/{l_3}$ 
for $\lambda_2^2=0.5,0.1,0.01$.}
\label{fig:eff_2x3M}
\end{figure}

The effective action at the 2-loop level is
\begin{equation}
\Gamma = n^{\frac{2}{3}} N^{\frac{4}{3}} \left[ - \frac{r_1 + r_2 + r_3 }{24 \lambda_2^2}  
+  \frac{1}{8} \int_0^4  \frac{dX  dL  dA}{r_1X+r_2L+r_3A} - \lambda_2^2 
\left( \frac{3}{2} f_3 + 4 f_4 \right) \right] .
\end{equation}
It is indeed of $O(N^{4\over 3})$ 
after choosing the 't Hooft coupling $\lambda_2$:
\begin{equation}
\lambda_2^2 = \frac{ n^{\frac{4}{3}} }{ f^4 N^{\frac{1}{3}} } .
\end{equation}
In the same way as in the previous section, we have estimated this effective action 
numerically. From Fig.\ref{fig:eff_2x3M} we can observe again that the fuzzy 
$S^2 \times S^2 \times S^2$ background is not stable. 
The same analysis in the previous section can be used here 
to determine into which configuration it tends to decay.
From such an analysis we conclude that the fuzzy $S^2 \times S^2 \times S^2 $ 
background decays toward fuzzy $S^2$. 
It is interesting to observe that $S^2 \times S^2 \times S^2$ becomes a local minimum of the
effective action at $t=1$
when the coupling $\lambda_2$ is strong enough.

\section{IIB matrix model analysis with respect to spins}
\setcounter{equation}{0}

After investigating bosonic and supersymmetric models with Myers term, 
we evaluate the effective action for IIB matrix model (\ref{action_IIB}) 
with the background (\ref{classical-sol}).
The results are
\footnote{The two loop amplitude $\Gamma_{2-loop}$ is evaluated in the Appendix with generic scale factors.}
\beqa
 \label{action2x3}
\Gamma_{tree} &\simeq&  n^{\frac{2}{3}} N^{\frac{4}{3}} \frac{1}{8} 
 \frac{ f^4 N^{\frac{1}{3}} }{ n^{\frac{4}{3}} }(r_1 + r_2 + r_3 ) , \n
\Gamma_{1-loop} &\simeq& O \left( N^{\frac{2}{3}} \right) , \n
\Gamma_{2-loop} &\simeq& n^{\frac{2}{3}} N^{\frac{4}{3}} 
\frac{  n^{\frac{4}{3}} }{ f^4 N^{\frac{1}{3}} } \frac{1}{2} f_3 .
\eeqa
We can explicitly check that the effective action scales as $N^{4\over 3}$
at the two loop level with the following choice of the 't Hoot coupling
\beqa
\Gamma &=&n^{\frac{2}{3}} N^{\frac{4}{3}} \left( \frac{1}{8 \lambda_2^2} 
 (r_1 + r_2 +r_3) + \frac{\lambda_2^2}{2} f_3 \right) ,  \n
\lambda_2^2 &=& \frac{ n^{\frac{4}{3}} }{ f^4 N^{\frac{1}{3}} } .
\eeqa
Since the 't Hoot coupling which is set by the overall scale of the background 
becomes a dynamical variable, we can minimize the effective action with respect to it
for fixed representations:
\begin{equation} \label{min-eff-2x3}
\Gamma_{min}= 2 \sqrt{\Gamma_{tree} \Gamma_{2-loop}}.
\end{equation}

\begin{figure}[htbp]
\epsfysize=8cm
\begin{center}
\vspace{1cm}
\hspace{0cm}
\epsfbox{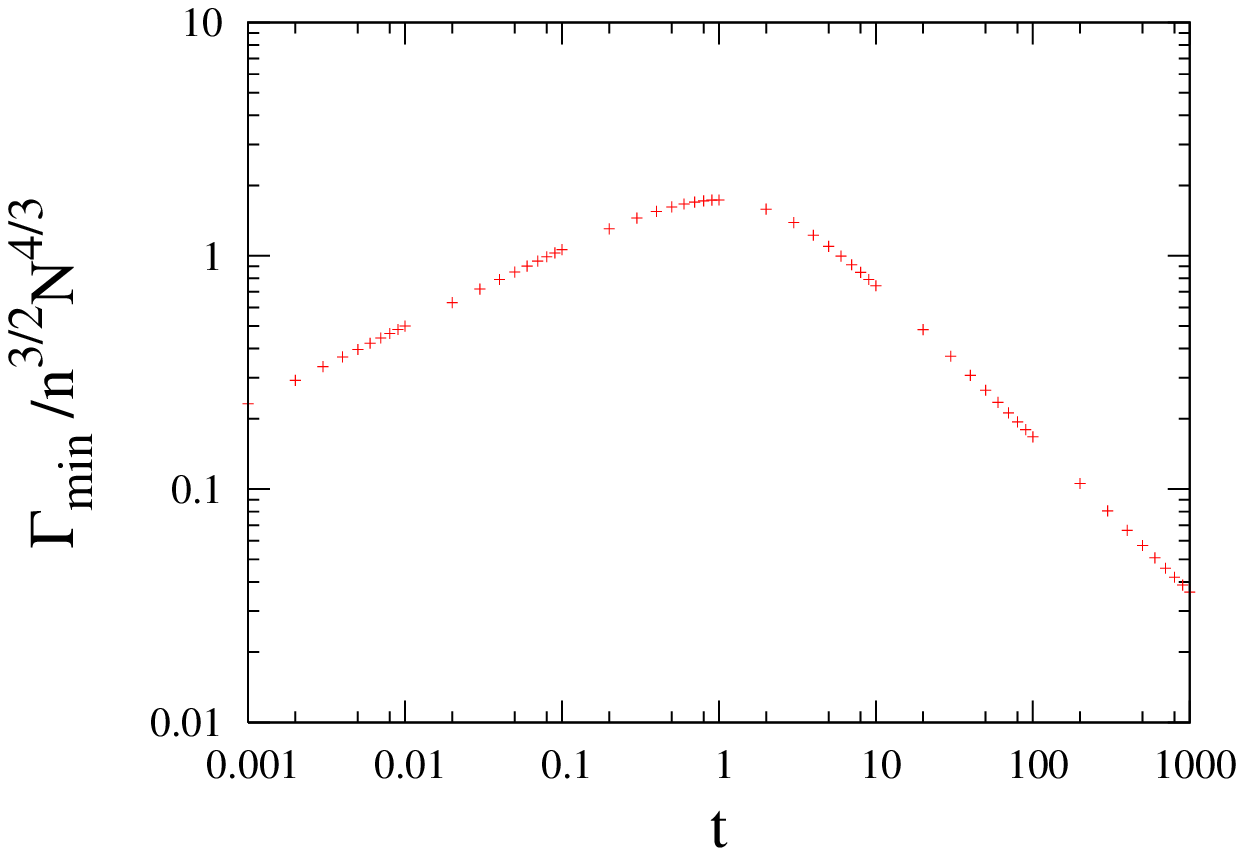}
\end{center}
\caption{${\Gamma_{min}}/{n^{3/2}N^{4/3}}$ against $t={l_1}/{l_3}={l_2}/{l_3}$.}
\label{fig:eff_2x3_6d.eps}
\end{figure}

We can now explore the minimum of the effective action 
(\ref{min-eff-2x3}) with respect to the representations $(l_1,l_2,l_3)$.
Fig.\ref{fig:eff_2x3_6d.eps} shows ${\Gamma_{min}}/{n^{\frac{3}{2}}N^{\frac{4}{3}}}$ 
against $t={l_1}/{l_3}={l_2}/{l_3}$. We note that
the action is decreasing faster in the $t>1$ region than $t<1$ region. We may thus conclude that
fuzzy $S^2 \times S^2 \times S^2 $ is not stable and it tends to decay toward fuzzy $S^2 
\times S^2$. This result is consistent with the previous investigations \cite{IKTT03,IT03}.

We can further demonstrate that the action (\ref{action2x3}) reduces to that of $S^2 \times S^2$ 
when we take a limit $l_1,l_2 \gg l_3$. 
It is because we can reexpress (\ref{action2x3}) as
\beqa
\Gamma''_{tree} &=& \frac{N}{8\lambda_3^2}\left( r + 1/r+R 
 \right) , \n
\Gamma''_{2-loop} &=& 2 N \lambda_3^2 l_3^2 \int^4_0 dXdYdZdLdMdNdAdBdC \n
&&\times \frac{W(X,Y,Z) W(L,M,N) W(A,B,C)}{(r X +r^{-1} L+ R A)(r Y +r^{-1} M+R B)(r Z +r^{-1} N+R C)} ,  \n
N &\simeq& 2l_1 2l_2 2l_3 \ , \ \lambda_3^2 = \frac{1}{f^4 2l_1 2l_2} \ , 
 \ R =\frac{l_3^2}{l_1 l_2} \ , \ r=\frac{l_1}{l_2} .
\eeqa
Here, we set $n=1$ for simplicity.

\begin{figure}[htbp]
\epsfysize=8cm
\begin{center}
\vspace{1cm}
\hspace{0cm}
\epsfbox{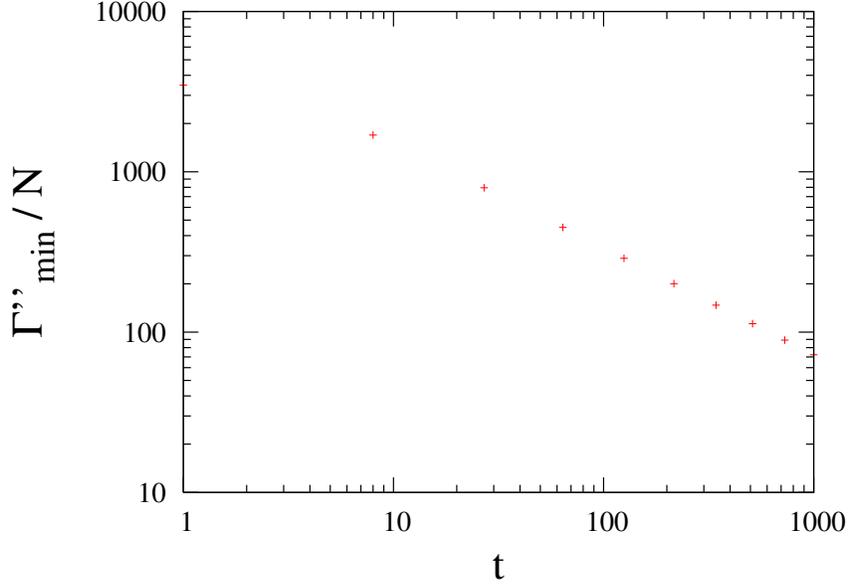}
\end{center}
\caption{${\Gamma''_{min}}/{N}$ against $t={l_1}/{l_3}={l_2}/{l_3}$.}
\label{fig:eff_2x3_4d.eps}
\end{figure}

Fig.\ref{fig:eff_2x3_4d.eps}  shows $ \Gamma''_{min} /N = 2 
\sqrt{ \Gamma''_{tree} \cdot \Gamma''_{2-loop} } / N $ 
against $t={l_1}/{l_3}={l_2}/{l_3}$ for $N=8\times 10^9$.  It demonstrates 
that the effective action scales in a 4d fashion as the background approaches $S^2 \times S^2$.

\section{Stability of fuzzy $S^2 \times S^2$ background}
\setcounter{equation}{0}

In this section, we investigate the stability of fuzzy $S^2 \times S^2$ background in detail
with respect to the scale factors  in addition to the change of the representations (spins).
Our investigation is motivated by an analogous study
for the fuzzy torus case \cite{BHKK04}.

We consider the background of the following type:
\beqa
\label{back-2x2}
 p_{\mu} &=& f_1 \left( \bar{j}_{\mu} \otimes 1  \right) \qquad  ( \mu = 1,2,3 ) , \n
 p_{\mu} &=& f_2 \left( 1 \otimes \hat{j}_{\mu} \right) \qquad  ( \mu = 4,5,6 ) , \n
 p_{\mu} &=& 0 \qquad ( \mu = 7,8,9,0 ) , \n
 \chi &=& 0 .
\eeqa
Here we set $n=1$ for simplicity. It generalizes the background in \cite{IT03} 
where the identical scale factor $f_1=f_2$ is assumed.

The effective actions are evaluated as
\footnote{It can again be read off from $\Gamma_{2-loop}$ in the Appendix.}
\beqa
\label{eff-2x2}
 {\Gamma'}_{tree} &=& \frac{N'}{8{\lambda'}^2} \left( r q^2 + \frac{1}{rq^2} \right) ,\n
 {\Gamma'}_{1-loop} &=& O (\log N') ,\n
 {\Gamma'}_{2-loop} &=& 4N' {\lambda'}^2 \int^4_0 dXdYdZdLdMdN \n
 &&\times \frac{(r q^2 X +\frac{1}{rq^2} L)}
 {(r q X +\frac{1}{rq} L)^2(r q Y +\frac{1}{rq} M)(rq Z +\frac{1}{rq} N)} \n
 &&\times W(X,Y,Z) W(L,M,N) ,
\eeqa
where
\begin{equation}
 {N'}=2l_1 2l_2  , \ {\lambda'}^2 = \frac{1}{f_1^2 f_2^2 N'} , \ 
 q = \frac{f_1}{f_2} , \ r = \frac{l_1}{l_2} .
\end{equation}

\begin{figure}[htbp]
\epsfysize=8cm
\begin{center}
\vspace{1cm}
\hspace{0cm}
\epsfbox{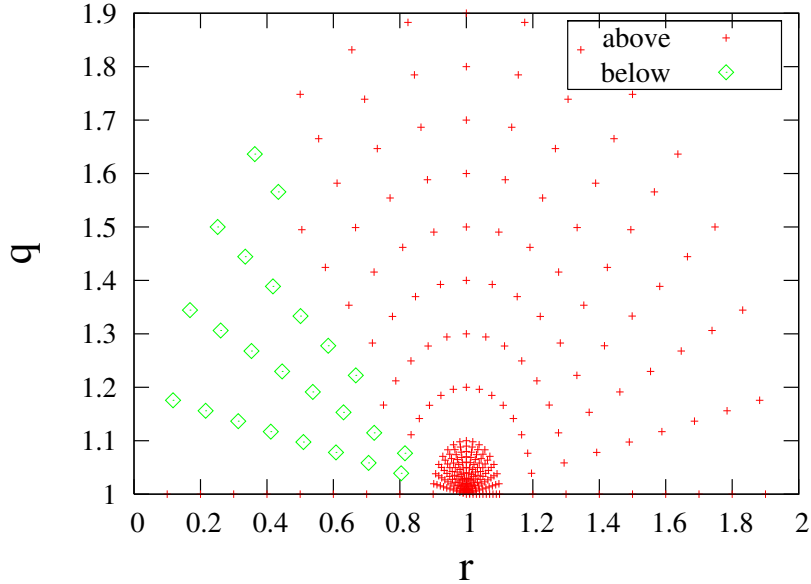}
\end{center}
\caption{3D plot:${\Gamma'_{min}}/{N'}$ against $r$ and $q$. $r-q$ section.}
\label{fig:eff_2x2_rq1.eps}
\end{figure}

\begin{figure}[htbp]
\epsfysize=8cm
\begin{center}
\vspace{1cm}
\hspace{0cm}
\epsfbox{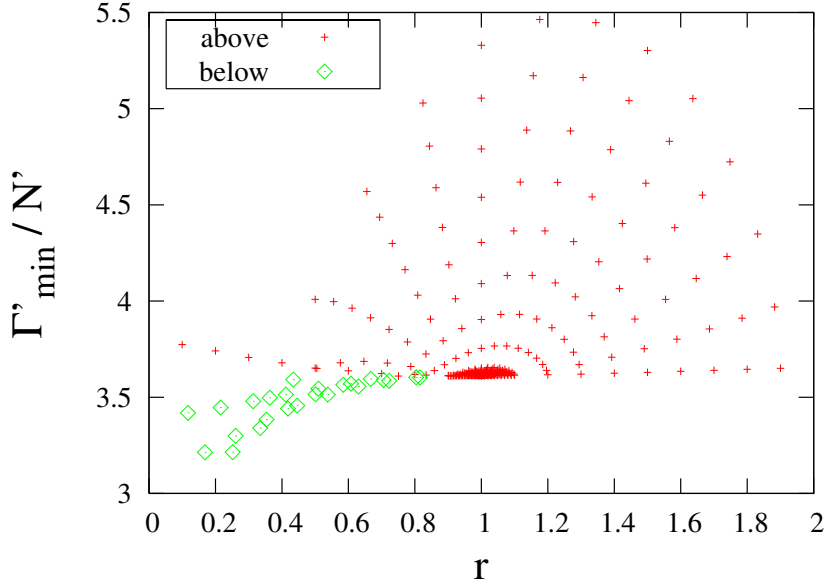}
\end{center}
\caption{3D plot:${\Gamma'_{min}}/{N'}$ against $r$ and $q$. $r-{\Gamma'_{min}}/{N'}$ section.}
\label{fig:eff_2x2_rq2.eps}
\end{figure}

Fig.\ref{fig:eff_2x2_rq1.eps} and Fig.\ref{fig:eff_2x2_rq2.eps} show  
$ \Gamma'_{min}/N' = 2 \sqrt{ \Gamma'_{tree} \cdot \Gamma'_{2-loop} } / N' $ 
against $r$ and $q$. 
The points represented by squares possess smaller effective actions
than that of the most symmetric point $r=q=1$. Furthermore
$\sqrt{rq}$ (the scale ratios of the two $S^2$) also decreases in this domain. 
Therefore we find that
fuzzy $S^2 \times S^2$ background is not stable when the both spins and scale 
factors are allowed to change at the two loop level.

\section{Local stability of the backgrounds}
\setcounter{equation}{0}

In the previous sections, we have investigated the stability of the backgrounds
globally by performing the multiple integrals numerically. 
In this section we also investigate the local stability of
the most symmetric background with respect to the small variations of the 
spins and scale factors.

In $S^2 \times S^2$ case, the minimum of the effective action (\ref{eff-2x2}) 
under the variations of $r=1+\delta ,q=1+\epsilon ,(\delta ,\epsilon \ll 1)$ becomes
\beqa
  {\Gamma'_{min}\over N' }&=& \frac{2 \sqrt{ \Gamma'_{tree} \cdot \Gamma'_{2-loop} 
  } }{N'} \equiv \frac{G}{4} , \n
  G &=& \left< \frac{(X+q^4r^2L)(1+q^4 
  r^2)r^2}{(X+q^2r^2L)^2(Y+q^2r^2M)(Z+q^2r^2N)} \right> \n
  &\simeq& 2 F' + ( -\frac{7}{2} F' + 6|C'|^2 + 6 |\varepsilon |^2 ) 
 \delta^2 \n
  &&{} \, + 2 (|C'|^2 +|\varepsilon' |^2) \epsilon^2 + (-4 F' 
 +8|C'|^2+8|\varepsilon' |^2) \delta \epsilon ,
\eeqa
where
\beqa
  F' &\equiv& \left< \frac{1}{(X+L)(Y+M)(Z+N)} \right> , \n
  |C'|^2 &\equiv& \left< \frac{L^2}{(X+L)^3(Y+M)(Z+N)} \right> , \n
  |\varepsilon'|^2 &\equiv& \left< \frac{LM}{(X+L)^2(Y+M)^2(Z+N)} \right> , \n
  \left< \ldots \right> &\equiv& \int_{0}^{4} dX dY dZ dL dM dN
  \ldots W(X,Y,Z) W(L,M,N).
 \eeqa
We can estimate $F',|C'|^2,|\varepsilon'|^2$ numerically as follows
\beqa
F' &=& 3.263930 \pm 0.19 \times 10^{-4} , \n
|C'|^2 &=& 1.056778 \pm 0.94 \times 10^{-5} , \n
|\varepsilon' |^2 &=& 0.8630541 \pm 0.68 \times 10^{-5} .
\eeqa
Let us consider the following ratio:
\begin{equation}
 \frac{G}{F'} = 2 + 0.0292 \delta^2 + 1.18 \epsilon^2 + 0.706 \epsilon \delta .
\end{equation}
To explore the stability of the background, we evaluate the eigenvalues of the quadratic forms 
in $\delta$ and $\epsilon$ above. 
The eigenvalues  are 
\begin{equation}
1.28,~~-0.07 \ .
\end{equation}
Since we find a negative eigenvalue, we conclude that $S^2 \times S^2$ is not stable 
in such a direction around the most symmetric point.

The same analysis can be applied to $S^2 \times S^2 \times S^2$. In this case, 
a relevant background corresponding to (\ref{back-2x2}) is
\beqa 
\label{back-2x3}
p_{\mu} &=& f_1 \left( \bar{j}_{\mu} \otimes 1 \otimes 1 \right) 
\otimes 1_n \qquad  ( \mu = 1,2,3 ) , \n
p_{\mu} &= &f_2 \left( 1 \otimes \hat{j}_{\mu} \otimes 1 \right) 
\otimes 1_n \qquad  ( \mu = 4,5,6 ) , \n
p_{\mu} &=& f_3 \left( 1 \otimes 1 \otimes \tilde{j}_{\mu} \right) 
\otimes 1_n \qquad  ( \mu = 7,8,9 ) , \n
p_{0} &=& 0 , \n
\chi &=& 0.
\eeqa
We evaluate the two loop effective action for this generic background
in the Appendix.
The minimum of the effective action with respect to the 't Hooft coupling is
\beqa
\Gamma_{min} &=& \frac{N^{4/3} H^{1/2}}{2^{1/2}} , \n
H &=& \left< \frac{(X+q_2^4r_2^2L+q_3^4r_3^2A)(1+q_2^4 r_2^2+q_3^4 r_3^2)
(r_2^2r_3^2)^{\frac{2}{3}}}{(X+q_2^4r_2^2L+q_3^4r_3^2A)^2(Y+q_2^2r_2^2M+
q_3^2r_3^2B)(Z+q_2^2r_2^2N+q_3^2r_3^2C)} \right> , \n
\left< \ldots \right> &\equiv& \int_{0}^{4} dX dY dZ dL dM dN dA dB dC \ldots \n
&&{} \qquad \times W(X,Y,Z) W(L,M,N) W(A,B,C) , \n
&&r_2=\frac{l_2}{l_1}, \ r_3=\frac{l_3}{l_1}, \ q_2=\frac{f_2}{f_1}, \ 
q_3=\frac{f_3}{f_1}.
\eeqa
The functional dependence of $H$ on $r_i=1+\delta_i , q_i=1+\epsilon_i $ is
\beqa
 H &=& 3F + (9|C|^2+9|\varepsilon|^2-\frac{8}{3}F)(\delta_2^2 + \delta_3^2) 
 -(\frac{17}{3}F-9|C|^2)\delta_2 \delta_3 \n
&&+ (9|C|^2+9|\varepsilon|^2 -\frac{3}{2}F)(\epsilon_2^2+\epsilon_3^2) + 
 (F-3|C|^2)\epsilon_2 \epsilon_3 \n
 &&+(12|C|^2+12|\varepsilon|^2-\frac{8}{3}F)(\delta_2 \epsilon_2 + \delta_3 
 \epsilon_3) \n
&&+(-\frac{3}{2}F-6|C|^2+12|\varepsilon|^2)(\delta_2 \epsilon_3 + \delta_3 \epsilon_2) ,
\eeqa
where
\beqa
  F &\equiv& \left< \frac{1}{(X+L+A)(Y+M+B)(Z+N+C)} \right> , \n
  |C|^2 &\equiv& \left< \frac{A^2}{(X+L+A)^3(Y+M+B)(Z+N+C)} \right> , \n
  |\varepsilon|^2 &\equiv& \left< \frac{AB}{(X+L+A)^2(Y+M+B)^2(Z+N+C)} \right> .
\eeqa
We can estimated $F,|C|^2,|\varepsilon|^2$  numerically as follows
\beqa
F &=& 3.993 \pm 0.25 \times 10^{-2} , \n
|C|^2 &=& 0.6032 \pm 0.46 \times 10^{-3} , \n
|\varepsilon |^2 &=& 0.4656 \pm 0.11 \times 10^{-2} .
\eeqa
It is again useful to consider the following ratio: 
\beqa
 \frac{H}{F} &=& 3 -0.26(\delta_2^2 + \delta_3^2) 
 -4.3\delta_2 \delta_3 \n
&&+ 0.91(\epsilon_2^2+\epsilon_3^2) - 
 0.55 \epsilon_2 \epsilon_3 \n
 &&+0.55(\delta_2 \epsilon_2 + \delta_3 
 \epsilon_3)
-1.0(\delta_2 \epsilon_3 + \delta_3 \epsilon_2) .
\eeqa
The eigenvalues of this quadratic form are
\beq
-2.43, ~~2.39, ~~0.69, ~~0.65 .
\eeq
The existence of the negative eigenvalue implies the instability of this background.
If we consider the $\delta_2,\delta_3$ subspace,
the eigenvalues are
\beq
-2.41, ~~1.89
\eeq
indicating the instability of this background
in agreement with section 5.

\section{Conclusions}

In this paper we have investigated the effective action of matrix models
with $S^2\times S^2\times S^2$ backgrounds
at the two loop level.
This class of 6 dimensional manifolds can be constructed
by using $SU(2)$ algebra which facilitates us to evaluate
the effective action.
We can change the size of each $S^2$ by choosing different
representations (spins) of $SU(2)$. Therefore we can probe manifolds
of different dimensionality such as $S^2$ (2d), $S^2\times S^2$ (4d)
and $S^2\times S^2\times S^2$ (6d) by collapsing some of $S^2$'s.
Since the background with the smallest effective action is most likely to be realized
in a particular model,
this investigation may shed light why our spacetime is 4 dimensional
in IIB matrix model context.

In the previous investigations, the large $N$ scaling behavior of the NC gauge theory
on these manifolds has been clarified.  In supersymmetric models,
the effective action scales as $N^2$, $N$ and
$N^{4\over 3}$ in 2, 4 and 6 dimensional manifolds respectively. 
It is always $O(N^2)$ and the one loop approximation is exact in bosonic models.
We have indeed verified this scaling behavior for 6 dimensional spacetime at the two loop level.
With the presence of Myers term, $S^2$ configuration  is favored since the effective
action can become negative.
On the other hand, 4 dimensional spacetime is favored in IIB matrix model
since the effective action is positive definite.

In accord with these expectations
we find that the fuzzy $S^2 \times S^2 \times S^2$ background 
is not stable when we vary the ratios of the spins 
in the following matrix models
\begin{itemize}
\item 10d bosonic matrix model with Myers term,
\item IIB matrix model with Myers term,
\item IIB matrix model.
\end{itemize}
In the first two matrix models with Myers term, 
$S^2 \times S^2 \times S^2$ tends to decay toward $S^2$.
 On the other hand
 $S^2 \times S^2 \times S^2$ tends to decay $S^2 \times S^2$ in the last one.
These results are consistent with the previous works \cite{IKTT03,IT03}.

We have further investigated the effective action around the symmetric $S^2\times S^2$
in detail. Under the change of
the ratios of the scale factors in addition to the spins, 
we find there are unstable directions of the $S^2 \times S^2$ background in IIB matrix model.

This instability does not imply that it eventually decays into $S^2$ since
the effective action is $O(N^2)$ in such a limit.
However this argument cannot be verified at the two loop level since
the effective action for $S^2$ behaves as follows up to the two loop level
\beq
aN^3f^4 + {b\over f^4N} ,
\eeq
where $a,b$ are $O(1)$.
It is therefore possible to make it $O(N)$ by choosing $1/f^4N^2$ to be $O(1)$
which is the same order with 4d manifolds.
However we argue that this is a two loop artifact since the $n$ loop contribution
can be estimated as $({1/ f^4N})^{n-1}$ and hence we need to adopt 
$1/f^4N$ as the 't Hooft coupling.
Nevertheless we cannot rule out the possible existence of instability
of a simple product space of $S^2\times S^2$.
In this respect a more symmetric 4d spacetime such as 
$CP^2=SU(3)/U(2)$ is interesting and it may not suffer from the instability
found here since $SU(3)$ symmetry permits only the over all
scale factor of the background \cite{Ki02, ABNNcp2}. 
It is also likely that a manifold with higher symmetry may lower
the effective action of IIB matrix model.

\begin{center} \begin{large}
Acknowledgments
\end{large} \end{center}
This work is supported in part by the Grant-in-Aid for Scientific
Research from the Ministry of Education, Science and Culture of Japan.
We thank  H. Aoki, T. Azuma, S. Bal , S. Iso, H. Kawai,  K. Nagao, J. Nishimura
and Y. Takayama for discussions.

\section*{Appendix A}
\renewcommand{\theequation}{A.\arabic{equation}}
\setcounter{equation}{0}

In this appendix, we evaluate the two loop effective action of IIB matrix model in the 
$S^2 \times S^2 \times S^2$ background (\ref{back-2x3}).
Our calculation which is an extension of those in \cite{IKTT03} incorporates the asymmetric 
scale factors.
The result in the $S^2 \times S^2$ background (\ref{back-2x2}) 
can be obtained by shrinking one of $S^2$'s.

We expand quantum fluctuations in terms of the tensor product of the matrix spherical harmonics:
\beqa
a^{\mu} &=& \sum_{jmpqst} a^{\mu}_{jmpqst} (Y_{jm} \otimes Y_{pq} \otimes Y_{st}) , \n
\varphi &=&\sum_{jmpqst} \varphi_{jmpqst} (Y_{jm} \otimes Y_{pq} \otimes Y_{st}) , \n
b &=& \sum_{jmpqst} b_{jmpqst} (Y_{jm} \otimes Y_{pq} \otimes Y_{st}) , \n
c &=& \sum_{jmpqst} c_{jmpqst} (Y_{jm} \otimes Y_{pq} \otimes Y_{st}) .
\eeqa
Here the sums of $j$, $p$ and $s$ run up to $2l_1$, $2l_2$ and $2l_3$ respectively. 
Then the propagators are derived from the kinetic terms of (\ref{q_action_IIB}):
\beqa
\left< a^{\mu} a^{\nu} \right> &=& \sum_{jmpqst}  
\left( P^2 \delta_{\mu \nu} + 2i f_{\mu \nu \rho} P^{\rho} \right)^{-1} 
(Y_{jm} \otimes Y_{pq} \otimes Y_{st})(Y_{jm}^{\dagger} \otimes 
Y_{pq}^{\dagger} \otimes Y_{st}^{\dagger}) , \n
\left< \varphi \bar{\varphi} \right> &=&  \sum_{jmpqst} 
\left( - \Gamma_{\mu} P_{\mu} \right)^{-1} 
(Y_{jm} \otimes Y_{pq} \otimes Y_{st})(Y_{jm}^{\dagger} 
\otimes Y_{pq}^{\dagger} \otimes Y_{st}^{\dagger}) , \n
<cb> &=& \sum_{jmpqst}\frac{1}{P^2} (Y_{jm} \otimes Y_{pq} 
\otimes Y_{st})(Y_{jm}^{\dagger} \otimes Y_{pq}^{\dagger} \otimes Y_{st}^{\dagger}) .
\eeqa
We exclude the singlet state $j=p=s=0$ in the propagators.
To calculate the leading contributions in the large $N$ limit, 
we expand the boson and the fermion propagators as
\beqa
\left( P^2 \delta_{\mu \nu} + 2i f_{\mu \nu \rho} P^{\rho} \right)^{-1} &\simeq& 
\frac{\delta_{\mu \nu}}{P^2} - 2i \frac{f_{\mu \nu \rho} P^{\rho}}{P^4} 
+ 4 \frac{I_{\mu \nu}(P)}{P^6} , \n
\left( - \Gamma_{\mu} P_{\mu} \right)^{-1} &\simeq& 
\frac{\Gamma^{\mu} P_{\mu} }{P^2} + 
\frac{i}{2}\frac{f_{\mu \nu \sigma}\Gamma^{\mu \nu \rho} P_{\sigma} P_{\rho} }{P^4}
- \frac{\Gamma \cdot f^2 \cdot P}{P^4} \n
&&+ \frac{P \cdot f^2 \cdot P P^{\mu} \Gamma_{\mu}}{P^6} .
\eeqa
We have introduced the following tensor
\beq
I_{\mu \nu} \equiv ( \bar{\delta}_{\mu \nu} \bar{P}^2 - \bar{P}_{\mu \nu} ) f_1^2 
+ ( \hat{\delta}_{\mu \nu} \hat{P}^2 - \hat{P}_{\mu \nu} ) f_2^2
+ ( \tilde{\delta}_{\mu \nu} \tilde{P}^2 - \tilde{P}_{\mu \nu} ) f_3^2 .
\eeq
The symbols $\bar{\quad}$, $\hat{\quad}$ and $\tilde{\quad}$ denote the sub-spaces 
whose Lorentz indices  $\mu$ run over $ (1,2,3) $, $ (4,5,6) $ and $ (7,8,9) $ respectively.
We also introduce 
\begin{equation}
P \cdot f^2 \cdot P \equiv f_1^2 \bar{P}^2 + f_2^2 \hat{P}^2 + f_3^2 \tilde{P}^2.
\end{equation}
Using these propagators, we can calculate the contributions to the two loop effective action 
from various interaction vertices as follows. 

4-gauge boson vertex is
\begin{equation}
V_4 = - \frac{1}{4} Tr [a_\mu ,a_\nu ]^2 .
\end{equation}
The leading contribution to the two loop effective action is
\beq
<-V_4> = 
\left< \frac{1}{P_1^2 P_2^2} \left\{ -45 + 6 \frac{P_3 \cdot f^2 \cdot P_3}{P_1^2 P_2^2} 
-12 \frac{P_2 \cdot f^2 \cdot P_2}{P_1^2 P_2^2}
-72 \frac{P_1 \cdot f^2 \cdot P_1}{P_1^4} \right\} \right>_P.
\eeq
We introduce the wave functions and averages as
\beqa
\Psi_{123} &\equiv& Tr (Y_{j_1m_1}Y_{j_2m_2}Y_{j_3m_3})
Tr (Y_{p_1q_1}Y_{p_2q_2}Y_{p_3q_3})Tr (Y_{s_1t_1}Y_{s_2t_2}Y_{s_3t_3}) , \n
\left< X \right>_P &\equiv& \sum_{j_i,p_i,s_i,m_i,q_i,t_i} \Psi_{123}^{*} X \Psi_{123} , \n
P_{i}^{\mu} ( Y_{j_im_i} Y_{p_iq_i} Y_{s_it_i} ) &\equiv&
 [p_{\mu},Y_{j_im_i} Y_{p_iq_i} Y_{s_it_i}] .
\eeqa
We define following functions:
\beqa
F_1 &=& \left< \frac{1}{P_1^4 P_2^4} \right>_P , \n
\tilde{g}_1 &=& \left< \frac{P_2 \cdot f^2 \cdot P_2}{P_1^4 P_2^4} \right>_P , \n
g_1 &=& \left< \frac{P_1 \cdot f^2 \cdot P_1}{P_1^6 P_2^2 } \right>_P , \n
g_2 &=& \left< \frac{P_3 \cdot f^2 \cdot P_3}{P_1^4 P_2^4} \right>_P .
\eeqa
Then
\begin{equation} \label{ap_4b}
<-V_4> = -45F_1 -12 \tilde{g}_1 -72 g_1 +6 g_2 .
\end{equation}

Ghost vertex is
\begin{equation}
V_g = Tr b \left[ p_{\mu} , \left[  a_{\mu} ,c \right] \right] .
\end{equation}
Their contribution is
\begin{equation} \label{ap_gh}
\frac{1}{2} < V_g V_g > = F_2 + 4 H_2 .
\end{equation}
Here
\beqa
F_2 &=& \left< \frac{P_2 \cdot P_3}{P_1^2 P_2^2 P_3^2} \right>_P , \n
H_2 &=& \left< \frac{P_2 \cdot I(1) \cdot P_3}{P_1^6 P_2^2 P_3^2} \right>_P ,
\eeqa
and
\begin{equation}
P_i \cdot I(j) \cdot P_k \equiv P_i^{\mu} I_{\mu \nu} (P_j) P_k^{\nu} .
\end{equation}

3-gauge boson vertex is
\begin{equation}
V_3 = - Tr P_{\mu} a_{\nu} \left[ a_{\mu}, a_{\nu} \right] .
\end{equation}
Their contribution is
\beqa
\label{ap_3b}
\frac{1}{2} < V_3 V_3 > &=& 9F_1 -9F_2 +12F_3 +8g'_1 -4 \tilde{g}'_1 + 2g_2 \n
&&+32H_1 - 36 H_2 -16H_3 +12H_4 -4H_5.
\eeqa
Newly introduced functions are defined as
\beqa
F_3 &=& \left< \frac{P_1 \cdot f^2 \cdot P_1}{P_1^4 P_2^2 P_3^2} \right>_P , \n
g'_1 &=& g_1-\frac{1}{N} \sum_{j,p,s} (2j+1)(2p+1)(2s+1) \n
&&\times \frac{f_1^4 j(j+1)+f_2^4p(p+1)+f_3^4s(s+1)}
{\left[ f_1^2j(j+1)+f_2^2p(p+1)+f_3^2s(s+1) \right]^4 }, \n
\tilde{g}'_1 &=& \tilde{g}_1 - \frac{1}{N} \sum_{j,p,s} (2j+1)(2p+1)(2s+1) \n
&&\times \frac{f_1^4 j(j+1)+f_2^4p(p+1)+f_3^4s(s+1)}
{\left[ f_1^2j(j+1)+f_2^2p(p+1)+f_3^2s(s+1) \right]^4 } , \n
H_1 &=& \left< \frac{P_1 \cdot I(2) \cdot P_1}{P_1^2 P_2^6 P_3^2} \right>_P , \n
H_3 &=& \left< \frac{P_2 \cdot I(1) \cdot P_3}{P_1^4 P_2^4 P_3^2} \right>_P , \n
H_4 &=& \left< \frac{P_1 \cdot I(2) \cdot P_1}{P_1^4 P_2^4 P_3^2} \right>_P , \n
H_5 &=& \left< \frac{P_2 \cdot I(1) \cdot P_3}{P_1^2 P_2^4 P_3^4} \right>_P .
\eeqa

Fermion vertex is
\begin{equation}
V_f=- \frac{1}{2} Tr \bar{\varphi} \Gamma_\mu \left[ a_\mu , \varphi \right] .
\end{equation}
Their contribution is
\beqa
\label{ap_fr}
\frac{1}{2} < V_f V_f > &=& -64F_2 +(-16 \tilde{g}'_1 +8g_2 +16F_3 +32H_4 ) \n
&&-32F_3 +64g'_1 +32 \tilde{g}'_1 -16 g_2 +64 H_2 +64H_3 .
\eeqa

After summing up (\ref{ap_4b}), (\ref{ap_gh}), (\ref{ap_3b}) and (\ref{ap_fr}), 
we find the 2-loop effective action:
\beqa
\Gamma_{2-loop} &=& 4F_3 + 32H_1 +32H_2+48H_3+(12+32)H_4 -4H_5 \n
&=& 4F_3 .
\eeqa
It is because
\beqa
H_1+H_2 &=& 0 , \n
H_3+H_4 &=& 0 , \n
H_3-H_5 &=& 0 .
\eeqa

The explicit form of $F_3$ is
\beqa
F_3 &=& \sum_{j_i,p_i,s_i} \n
&&\times \frac{(2j_1+1)(2p_1+1)(2s_1+1) 
\left[f_1^4 j_1(j_1+1)+f_2^4 p_1(p_1+1)+f_3^4 s_1(s_1+1) \right] }
{\left[ f_1^2 j_1(j_1+1)+f_2^2 p_1(p_1+1)+f_3^2 s_1(s_1+1)\right]^2} \n
&&\times \frac{(2j_2+1)(2p_2+1)(2s_2+1)}
{\left[ f_1^2 j_2(j_2+1)+f_2^2 p_2(p_2+1)+f_3^2 s_2(s_2+1)\right]} \n
&&\times \frac{(2j_3+1)(2p_3+1)(2s_3+1)}
{\left[ f_1^2 j_3(j_3+1)+f_2^2 p_3(p_3+1)+f_3^2 s_3(s_3+1)\right]} \n
&&\times \left\{ \begin{array}{ccc} j_1 & j_2 & j_3 \\ l_1 & l_1 & l_1 \end{array} \right\}^2 
\left\{ \begin{array}{ccc} p_1 & p_2 & p_3 \\ l_2 & l_2 & l_2 \end{array} \right\}^2
\left\{ \begin{array}{ccc} s_1 & s_2 & s_3 \\ l_3 & l_3 & l_3 \end{array} \right\}^2 .
\eeqa
In the large $N$ limit, we can use the following approximations
\beqa
j_i(j_i+1) &\rightarrow& j_i^2 , \n
2j_i+1 &\rightarrow& 2j_i , \n
\sum_{j_i=1}^{2l_1} &\rightarrow& \int_{0}^{2l_1} dj_i , \n
l_1^3 \left\{ \begin{array}{ccc} j_1 & j_2 & j_3 \\ l_1 & l_1 & l_1 \end{array} \right\}^2 
&\rightarrow& W(j_1^2,j_2^2,j_3^2).
\eeqa
We also define new variables as
\beqa
&&X = j_1^2 , \ Y = j_2^2 , \ Z = j_3^2 , \n
&&L = p_1^2 , \ M = p_2^2 , \ N = p_3^2 , \n
&&A = s_1^2 , \ B = s_2^2 , \ C = s_3^2 .
\eeqa
Finally $F_3$ assumes the following expression
\beqa
F_3 &=& \frac{l_1 l_2 l_3}{(f_1f_2f_3)^{4/3}} \int^4_0 dXdYdZdLdMdNdAdBdC \n
&&\times \frac{q_1^2 r_1 X + q_2^2 r_2 L+ q_3^2 r_3 A}{(q_1 r_1 X + q_2 r_2 L+ q_3 r_3 A)^2} \n
&&\times \frac{1}{( q_1 r_1 Y + q_2 r_2 M+ q_3 r_3 B)(q_1 r_1 Z + q_2 r_2 N+ q_3 r_3 C)} \n
&&\times  W(X,Y,Z) W(L,M,N) W(A,B,C) ,
\eeqa
where
\beqa
&&r_1=\left( \frac{l_1 l_1}{l_2 l_3} \right)^{\frac{2}{3}} \ , \ 
r_2=\left( \frac{l_2 l_2}{l_1 l_3} \right)^{\frac{2}{3}} \ , \ 
r_3=\left( \frac{l_3 l_3}{l_1 l_2} \right)^{\frac{2}{3}} , \n
&&q_1=\left( \frac{f_1 f_1}{f_2 f_3} \right)^{\frac{2}{3}} \ , \ 
q_2=\left( \frac{f_2 f_2}{f_1 f_3} \right)^{\frac{2}{3}} \ , \ 
q_3=\left( \frac{f_3 f_3}{f_1 f_2} \right)^{\frac{2}{3}} .
\eeqa

\end{document}